\documentclass[11pt]{article}
\pdfoutput=1
\usepackage{jheppub}
\usepackage[T1]{fontenc}

\usepackage{verbatim}

\def\beq{\begin{equation}}
\def\eeq{\end{equation}}
\def\bec{\begin{center}}
\def\eec{\end{center}}

\newcommand{\bea}{\begin{eqnarray}}
\newcommand{\eea}{\end{eqnarray}}
\newcommand{\non}{\nonumber}
\newcommand{\mc}{\mathcal}

\newcommand{\mr}{\mathrm}

\def\muf{\mu_\mr{F}}
\def\mur{\mu_\mr{R}}

\def\({\left(} 
\def\){\right)} 

\def\neu{\tilde{\chi}}
\def\cha{\tilde{\chi}}
\def\usqr{\tilde{u}_{R}}
\def\dsqr{\tilde{d}_{R}}
\def\usql{\tilde{u}_{L}}
\def\dsql{\tilde{d}_{L}}
\def\csqr{\tilde{c}_{R}}
\def\ssqr{\tilde{s}_{R}}
\def\csql{\tilde{c}_{L}}
\def\ssql{\tilde{s}_{L}}
\def\tsqr{\tilde{t}_{2}}
\def\bsqr{\tilde{b}_{2}}
\def\tsql{\tilde{t}_{1}}
\def\bsql{\tilde{b}_{1}}
\def\jet{j}

\newcommand{\GeV}{{\mathrm{GeV}}}
\newcommand{\TeV}{{\mathrm{TeV}}}
\newcommand{\fb}{{\mathrm{fb}}}

\def\PROSPINO{{\tt PROSPINO}}
\newcommand{\RESUMMINO}{{\tt RESUMMINO}}

\newcommand{\madgraph}{{\tt MadGraph}}
\newcommand{\POWHEG}{{\tt POWHEG}}
\newcommand{\POWHEGBOX}{{\tt POWHEG-BOX}}
\newcommand{\PYTHIA}{{\tt PYTHIA}}
\newcommand{\feynarts}{{\tt FeynArts}}
\newcommand{\formcalc}{{\tt FormCalc}}

\newcommand{\NLOPS}{{\tt NLO+PS}}

\title{\bf 
Precise predictions for electroweakino-pair production in association with a jet at the LHC}

\author{Julien Baglio$^{1,\, 2,\, 3}_{}$,}
\author{Barbara J\"ager$^1_{}$,}
\author{and Matthias Kesenheimer$^1_{}$}
\affiliation{$^1_{}$ Institute for Theoretical Physics,  University of
  T\"ubingen, Auf der Morgenstelle 14, 72076~T\"ubingen, Germany}
\affiliation{$^2_{}$ Institute for Advanced Study, Durham
  University, Cosin's Hall, Palace Green, Durham DH1 3RL,
  United~Kingdom.}
\affiliation{$^3_{}$ Institute for Particle Physics Phenomenology,
  Department of Physics, Durham University, South Road, Durham DH1
  3LE, United~Kingdom.}
\emailAdd{julien.baglio@uni-tuebingen.de}
\emailAdd{barbara.jaeger@itp.uni-tuebingen.de}
\emailAdd{matthias.kesenheimer@uni-tuebingen.de}

\abstract{
We present the full NLO SUSY-QCD corrections to the pair production of neutralinos 
and charginos  at the LHC in association with a jet and
their matching to parton-shower programs in the framework of the
\POWHEGBOX{} package. The code we have developed provides a SUSY Les
Houches Accord interface for setting electroweak and supersymmetric input
parameters. Decays of the neutralinos and charginos and parton-shower
effects can be simulated with multi-purpose programs such as \PYTHIA. The capabilities of the code
are illustrated by phenomenological results for a parameter point in
the framework of pMSSM10, compatible with present experimental limits
on supersymmetry. We find that NLO-QCD corrections as well as
parton-shower effects are of primary importance for the appropriate description of jet distributions. 
}

\arxivnumber{1711.00730}
\keywords{Supersymmetry Phenomenology, NLO Computations}

\begin{document}

\thispagestyle{empty}
\def\thefootnote{\fnsymbol{footnote}}
\setcounter{footnote}{1}

\setcounter{page}{0}
\maketitle
\flushbottom

\def\thefootnote{\arabic{footnote}}
\setcounter{footnote}{0}
%
%
\section{Introduction}

After a very successful Run~I of the CERN Large Hadron Collider (LHC)
at energies of 7 and 8~TeV in which, notably, the ATLAS and CMS
collaborations discovered a particle with properties compatible with a
Standard Model (SM) Higgs boson~\cite{Aad:2012tfa,Chatrchyan:2012xdj},
LHC Run~II at 13~TeV started in 2015. This unprecedented energy allows to probe
an uncharted territory that may conceal new elementary
particles. Even though the observation of a Higgs boson has advanced
our understanding of the electroweak symmetry breaking mechanism,
important questions remain open that are pointing towards the need for
physics beyond the SM (BSM). In particular, astrophysical and
cosmological observations strongly imply the existence of Dark Matter
(DM), a new type of matter that interacts only weakly with SM
particles. The SM does not contain any elementary particles that are 
suitable candidates for DM. Amongst the different BSM scenarios on
the market, supersymmetry (SUSY) offers a very promising avenue to
account for DM (see, e.g., Ref.~\cite{Abercrombie:2015wmb} for a
recent review). SUSY features new particles that differ from their SM
counterparts by their spin and acquire large masses by the mechanism
of SUSY breaking. In the minimal supersymmetric extension of the SM
(the MSSM), the conservation of a discrete symmetry called R-parity
ensures that the lightest SUSY particle (LSP) is stable. In many SUSY
scenarios this particle is the lightest neutralino, a mixture of the
superpartners of the Higgs, the photon, and the $Z$~bosons. Being stable and
electrically neutral, it provides an excellent candidate for fermionic
DM. In the following, we will refer to neutralinos and charginos
generically as electroweakinos or simply weakinos. 

A very clean signature of DM at the LHC would be provided by a
significant amount of missing transverse energy in association with a
tagging jet or lepton, so-called mono-jet or mono-lepton
signatures. These can be produced by LSP pairs in two ways:
Either by the production of heavier SUSY particles that decay into
LSPs and additional leptons or jets; or by the direct production of 
a pair of LSPs in association with a jet.
Weakino-pair production is a pure electroweak  (EW) process at lowest
order. As a consequence, the associated production cross sections are small, and
the respective exclusion limits on the MSSM parameters obtained by
ATLAS and CMS are less severe than those obtained by the QCD mediated
squark and gluino production processes with significantly larger cross
sections, see
e.g. Refs.~\cite{Khachatryan:2014qwa,Aad:2015jqa,Aad:2015eda,CMS:2016saj}
at 8~TeV and Refs.~\cite{ATLAS:2017ocr,ATLAS:2017uun,CMS:2017sqn} at
13~TeV.  Thus, direct production of weakino pairs plus one identified
jet resulting in mono-jet signatures is of particular phenomenological
relevance.

The direct production of weakino pairs without an additional jet at
the lowest order has been studied for decades. The first calculation
of the next-to-leading-order (NLO) SUSY-QCD corrections to the total
cross section at hadron colliders was presented in
Ref.~\cite{Beenakker:1999xh}. The public computer program
\PROSPINO{}~\cite{Beenakker:1996ed} contains the NLO SUSY-QCD
corrections for the total cross sections. Resummation effects, in
particular for the transverse momentum distributions, were provided in
Ref.~\cite{Debove:2009ia}. That work showed that leading-order (LO)
calculations are not appropriate for predicting transverse momentum
spectra. Threshold corrections were provided in
Refs.~\cite{Li:2007ih,Debove:2010kf} and the combination of the latter
with transverse-momentum resummation effects was provided in
Ref.~\cite{Debove:2011xj} and implemented in the public package
\RESUMMINO{}~\cite{Fuks:2013vua}. As far as EW corrections are
concerned, they were calculated in Ref.~\cite{Hao:2006df} for the
associated production of a chargino and a neutralino at the LHC and
found to be moderate for representative parameter points. We provided
an interface of the NLO SUSY-QCD calculation for weakino-pair
production to parton shower programs~\cite{Baglio:2016rjx} via the
\POWHEG{} matching procedure~\cite{Nason:2004rx,Frixione:2007vw},
allowing for simulations including parton-shower, underlying-event,
and multi-parton interaction effects using Monte-Carlo programs such
as \PYTHIA{}~\cite{Sjostrand:2006za}.

Predictions for weakino-pair production in association with an
additional jet are less advanced. The LO differential cross sections
for all types of weakino pairs plus an extra hard jet were available
in Ref.~\cite{Giudice:2010wb} (and references therein), while the NLO
SUSY-QCD corrections for the neutralino-pair production mode in
association with a hard jet were calculated in
Ref.~\cite{Cullen:2012eh}. The latter calculation, however, does not
generically account for the complicated structure of the on-shell
resonances that may appear in the real-emission contributions of such
processes. Moreover, no dedicated interface of an NLO-QCD calculation
to parton-shower programs exists yet.

In this paper we provide that missing piece building on experience
gained in the implementation of weakino-pair production at the
LHC~\cite{Baglio:2016rjx}. We present a code package for all weakino-pair 
production processes (i.e.\ neutralino-neutralino,
neutralino-chargino and chargino-chargino)  in association with an
identified jet at the LHC, including the full NLO SUSY-QCD corrections
to cross sections and differential distributions within arbitrary
experimental selection cuts, matched to parton shower programs via the
\POWHEG{} method. We use the framework of the
\POWHEGBOX{}~\cite{Alioli:2010xd}, a public repository for the
simulation of scattering processes at hadron colliders at NLO-QCD
accuracy matched with parton shower programs. We provide, for the
first time, the subtraction of the on-shell resonance structure of the
real-emission corrections to a process with three 
particles in the final state at the lowest order. We find that the 
SUSY-QCD corrections modify the differential cross sections by up to $15$\% in certain phase-space regions at the LHC, in a way that cannot be accounted for with a constant $K$--factor.

In the following section, we briefly describe the technical aspects of
our calculation that are specific to the implementation of weakino-pair 
production processes in the context of the \POWHEGBOX, in particular the 
subtraction of the on-shell resonances. In Sec.~\ref{sec:pheno} we provide 
representative numerical results using a parameter point in the framework of 
the pMSSM10~\cite{deVries:2015hva}, compatible with present experimental
limits on SUSY. This illustrates the importance of NLO SUSY-QCD and
parton-shower effects in the correct description of the jet activity
in DM mono-jet searches at the LHC. Our conclusions are given in
Sec.~\ref{sec:conc}.

%
\section{Framework of the calculation}
\label{sec:calc}

Our calculation of the production of a pair of weakinos in association
with a jet builds on the experience we have gained in our previous
implementation of weakino-pair production processes in the framework
of the \POWHEGBOX{}~\cite{Baglio:2016rjx}. In this section, we present
the specific aspects of our implementation and, in particular, extensions 
with respect to the calculation of weakino-pair production.

\begin{figure}[t]
\bec
\includegraphics[width=0.9\textwidth,clip]{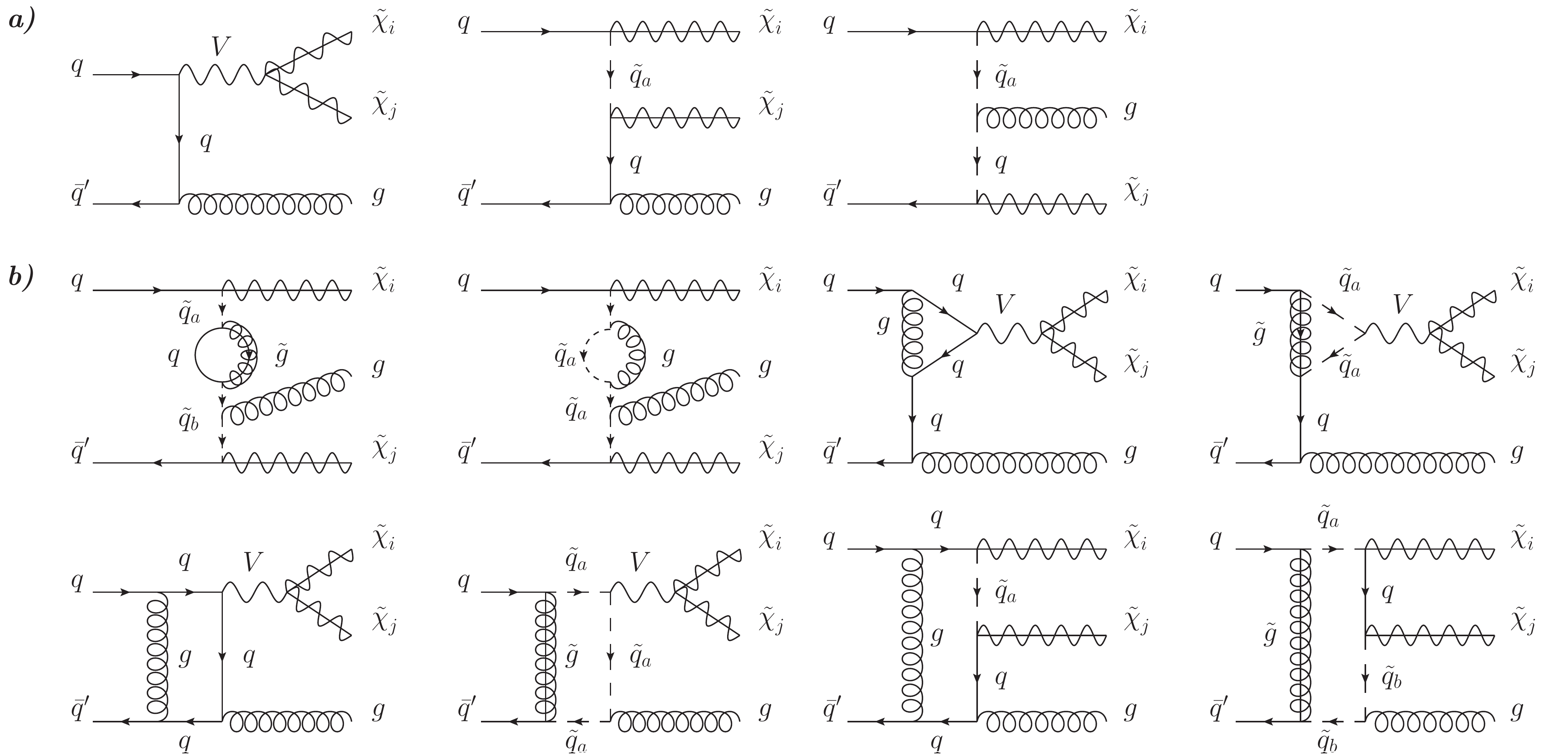}
\caption{Representative (a) tree-level and (b) one-loop diagrams for
  the production of a pair of weakinos, $\neu_i \neu_j$, 
  with a jet. Depending on the
  types $i,j$ of the produced weakinos, $V$ stands for $W_{}^\pm/Z/\gamma$,
  and $a,b=1,2$.}
\label{fig:virt_diagrams}
\eec
\end{figure}

At the lowest order the production of a pair of weakinos in
association with a jet proceeds via three parton-level channels:
$q\bar{q}'\to \neu_i^{}\neu_j^{} g$, $q g\to \neu_i^{}\neu_j^{} q$,
and $\bar{q} g\to \neu_i^{}\neu_j^{} \bar{q}$, where $\neu_i^{}$
stands for either a neutralino $\neu_i^0$ ($i=1\cdots 4$), or a
chargino $\neu_i^{} = \neu_i^\pm$ ($i=1,2$). Some representative
diagrams for the $q\bar{q}'$-induced partonic channels are depicted in
Fig.~\ref{fig:virt_diagrams}~a). 
We assume four massless active quark flavors in the running
of the strong coupling constant and take finite bottom-quark and
top-quark mass effects emerging in loop diagrams into account in the
calculation. The Cabibbo-Kobayashi-Maskawa matrix is taken diagonal
and we also use four-flavor-scheme parton distribution functions
with no bottom-quark initial states.

In the gluon-induced channels $q(\bar{q}) g\to \neu_i^{}\neu_j^{} q
(\bar{q})$, single resonances can already occur at the lowest order
when an intermediate squark happens to be on-shell in an $s$-channel
diagram. These resonance effects are regulated by using the
complex-mass scheme~\cite{Denner:2006ic}. We replace the mass of the
squark, $m_{\tilde{q}_k^{}}$, with a complex mass
$\mu_{\tilde{q}_k^{}}$ such that $\mu_{\tilde{q}_k^{}}^2 =
m_{\tilde{q}_k^{}}^2 - i \Gamma_{\tilde{q}_k^{}} m_{\tilde{q}_k^{}}$,
where  $\Gamma_{\tilde{q}_k^{}}$ denotes the physical decay width of
the squark. We stress that this procedure is gauge invariant.

\subsection{Virtual corrections}

The virtual corrections to the three parton-level processes
$q\bar{q}'\to \neu_i^{}\neu_j^{} g$, $q g\to \neu_i^{}\neu_j^{} q$,
and $\bar{q} g\to \neu_i^{}\neu_j^{} \bar{q}$ are classified as
self-energy, triangle, box, and pentagons corrections, with gluon,
gluino, quark, or squark exchange, see Fig~\ref{fig:virt_diagrams}~b)
for some representative diagrams. In our calculation, diagrams
containing a neutral Higgs boson coupling to top-quark and squark
loops and decaying to a pair of weakinos are included.
Resonance effects of these Higgs particles in $s$-channel 
diagrams are regulated by a finite Higgs width. We have used {\tt 
FeynArts 3.9}~\cite{Hahn:2000kx} to generate the virtual diagrams
and {\tt FormCalc 9.4}~\cite{Hahn:1998yk} to calculate, in the
Feynman-'t Hooft gauge, the amplitudes using the MSSM-CT model file of
Ref.~\cite{Fritzsche:2013fta}. In order to regularize single
resonances that we encounter already at the Born level, the
complex-mass scheme~\cite{Denner:2006ic} for squark and gluino masses
is used. Scalar and tensor one-loop integrals are numerically
evaluated with the computer package {\tt
  COLLIER-1.0}~\cite{Denner:2016kdg}.

For the regularization of ultraviolet (UV) divergences 
we follow the procedure of our previous calculation and use the
dimensional regularization scheme (DREG), the standard procedure of
the \POWHEGBOX{}. The entire calculation is done in $D=4-2\varepsilon$
dimensions, which is known to break supersymmetry at the level of the
gauge interactions by introducing a mismatch in the $(D-2)$ transverse
degrees of freedom of the gauge bosons and the two degrees of freedom
of the gauginos. In particular, this means that the
quark-squark-weakino Yukawa coupling $\hat{g}$ and the associated
$SU(2)$ gauge coupling $g$ are no longer equal to all orders in the
perturbative expansion. To restore this symmetry, we introduce a
finite SUSY restoring counter-term at NLO in the strong coupling
$\alpha_s$~\cite{Martin:1993yx,Beenakker:1999xh,Hollik:2001cz},
\begin{align}
\hat{g} = g \left( 1 - \frac{\alpha_s^{}}{6\pi}\right).
\label{eq:susyrestore}
\end{align}
The expansion in $\alpha_s$ is done consistently to retain only the
$\mathcal{O}(\alpha_s)$ term that is induced by this finite SUSY
restoring counter-term in the amplitude squared.

In order to cancel the UV divergences we perform a renormalization
procedure and calculate the suitable counter-terms necessary to define
finite physical input parameters. We use the on-shell scheme for the
renormalization of the wave functions of external colored particles
(quarks, gluons), as well as for the squark, gluino, and top masses
emerging in internal propagators. The strong coupling constant
$\alpha_s^{}$, however, is renormalized in the $\overline{\rm MS}$
scheme. In particular, the renormalization constant of the strong
coupling constant, $\delta Z_{g_{s}}$, is calculated using the gluon
field renormalization constant $\delta Z_{gg}^{}$ provided by {\tt
  FormCalc} (note that $\delta Z_{gg}^{}$ is in principle a
gauge-dependent renormalization constant and is here calculated in the
Feynman-'t Hooft gauge). We thus have
\begin{align}
\delta Z_{g_{s}}^{} = - \frac{3}{2}\delta Z_{gg}^{}|_{\rm div}^{} +
  \delta Z_{3g}^{} + \delta Z_{\rm log}^{},
\label{dzgs1}
\end{align}
with $\delta Z_{gg}^{}|_{\rm div}^{} $ being the divergent part of the
gluon field renormalization constant,
\begin{align}
Z_{gg}^{}|_{\rm div}^{} =
-\frac{3}{4\pi} \alpha_s^{} \Delta,
\label{dzgs2}
\end{align}
and
\begin{align}
\delta Z_{3g}^{} = -\frac{3}{2\pi} \alpha_s^{}\Delta ,
\label{dzgs3}
\end{align}
in the MSSM, with $\Delta = 1/\varepsilon-\gamma_E^{}+\ln
4\pi$ in the $\overline{\rm MS}$ scheme. The quantity $\gamma_E^{}$
is the Euler-Mascheroni constant. The last term in Eq.~(\ref{dzgs1})
is an additional finite shift to account for the decoupling of heavy
particles in the running of the strong coupling constant
$\alpha_s^{}$. The decoupling of heavy particles has first been discussed in Ref.~\cite{Collins:1978xx} and can be understood as follows.

The renormalization counterterm for the strong coupling constant can
be associated with the $\beta$-function, defined by the renormalization
group equation of QCD,
\begin{align}
\mu_R^2 \frac{d}{d\mu_R^2} \alpha_s = \beta(\alpha_s) = -\alpha_s^2
  \sum_{n} \beta_n \alpha_s^n,
\end{align}
where $\mu_R$ is the renormalization scale.
At the one-loop order we have (see, for example, 
Ref.~\cite{Beenakker:1996ch}) $\delta Z_{g_{s}}^{} = -\alpha_s \Delta
\beta_0/2$. With Eqs.~(\ref{dzgs1})--(\ref{dzgs3}), this means
$\beta_0^{} = 3/4\pi$, and all particles, including the heavy states,
are contributing to the running of $\alpha_s^{}$, leading to potentially
large logarithms, if the scale $\mu_R^{}$ is significantly different from the
masses of the various heavy states of  the theory. In the SM with only four active
flavors contributing to the running of $\alpha_s^{}$, we have
$\beta_0^{\rm light} = 25/12\pi$. Following
Ref.~\cite{Beenakker:1996ch}, the heavier states can
be subtracted by rescaling $\alpha_s^{}$ so that only the light four
active quark flavors contribute to the running of $\alpha_s^{}$ as in
the SM, with the bottom and top quarks taken as massive, decoupled
particles. This can be achieved in our calculation by adding the finite term $\delta
Z_{\rm log}^{}$ to the strong coupling constant counterterm, to subtract
the logarithms that arise from the masses of squarks
($m_{\tilde{q_i^{}}}^{}$), gluino ($m_{\tilde{g}}^{}$), top quark
($m_t^{}$) and bottom quark ($m_b^{}$),
\begin{align}
  \delta Z_{\rm log}^{} =
  -\frac{\alpha_s}{8\pi}\left[2\ln\frac{m_{\tilde{g}}^2}{\mu_R^2}+\frac{1}{6}
  \sum\limits_{i=1}^{12}\ln\frac{m_{\tilde{q_i}}^2}{\mu_R^2}+\frac{2}{3}
  \ln\frac{m_{t}^2}{\mu_R^2}+\frac{2}{3}\ln\frac{m_{b}^2}{\mu_R^2}\right].
 \end{align}
With this shift the one-loop running of $\alpha_s$ with
four light degrees of freedom is recovered:
\begin{align}
  \mu_R^2 \frac{d}{d\mu_R^2} \alpha_s
  &
    = -\alpha_s \left[\beta_0^{} -
    \frac{1}{4\pi}\left(-2-2-\frac{2}{3}-\frac{2}{3}\right)\right] 
    \nonumber\\
  & = -\alpha_s \beta_0^{\rm light}.
\end{align}
Other fundamental parameters emerging in our calculation, such as the
electroweak coupling constant, do not require renormalization at NLO
in QCD.

\subsection{Real emission corrections}

\begin{figure}[t]
\bec
\includegraphics[width=0.9\textwidth,clip]{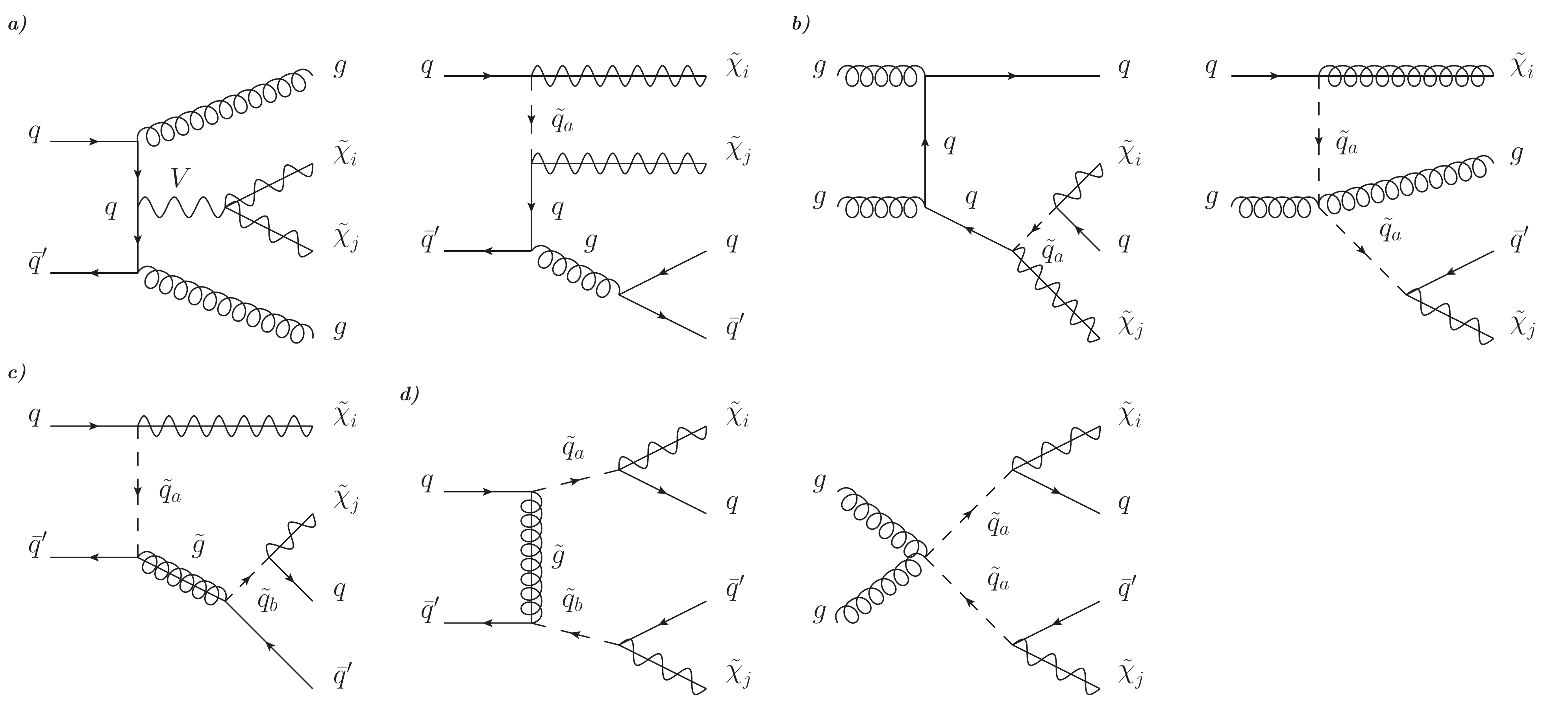}
\caption{Representative (a) non-resonant, (b) squark singly-resonant,
  (c) gluino singly-resonant, and (d) squark doubly resonant
  real-emission diagrams for the production of a pair of weakinos with
  a jet.}
\label{fig:real_diagrams}
\eec
\end{figure}

The calculation of the real-emission corrections to weakino-pair
production in association with a jet requires the evaluation of all
partonic subprocesses contributing to the reaction $p p\to \neu_i
\neu_j \jet \jet$. This includes quark--anti-quark annihilation
processes with two gluons in the final state and all possible crossed
modes, as well as  the additional class of processes with two
(anti-)quarks in the initial and in the final state.
Matrix elements for all of these subprocesses can in principle be
generated with the help of automated tools, such as the generator
provided by the \POWHEGBOX{} based on \madgraph{}, or the
\feynarts{}/\formcalc{} packages.
However, the perturbatively meaningful and numerically stable
evaluation of these contributions requires the design of a
well-defined procedure for the treatment of single and  double
on-shell resonances that are associated with Born processes different
from $p p\to \neu_i \neu_j \jet$, and should thus not be considered as
real-emission corrections to that reaction. In particular, this
includes the associated production of gluinos or squarks that
subsequently decay into a weakino and a parton. Examples of diagrams
that can become singly and or doubly resonant in specific regions of
the SUSY parameter space are shown in
Fig.~\ref{fig:real_diagrams}~(b,c) and
Fig.~\ref{fig:real_diagrams}~(d), respectively.

Note that single on-shell resonances emerge not only in weakino-pair plus jet production, but also in other supersymmetric production processes involving squarks or gluinos~\cite{Beenakker:1996ch,GoncalvesNetto:2012yt,Gavin:2013kga,Gavin:2014yga,Baglio:2016rjx}. The necessity of the subtraction method we are referring to has first been described in Ref.~\cite{Beenakker:1996ch}. Technical details can also be found in Ref.~\cite{GoncalvesNetto:2012yt}.
For the case of weakino-pair plus jet production the subtraction procedure of single resonances developed in Ref.~\cite{Baglio:2016rjx} can be used without
significant changes. However, the subtraction of double resonances requires a more sophisticated scheme.
\begin{figure}[t]
\bec
\includegraphics[width=0.35\textwidth]{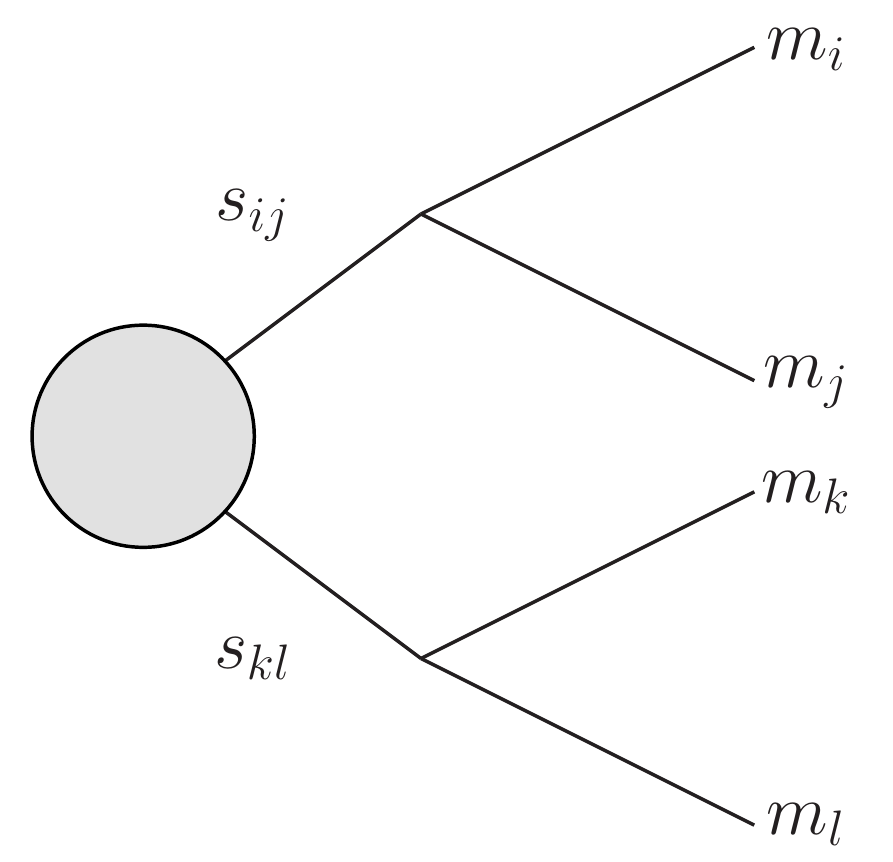}
\caption{Final-state topology of a doubly resonant diagram for the production of
  the particle pairs ($i,j$) and ($k,l$).}
\label{fig:double-resonance}
\eec
\end{figure}
The doubly resonant part of a matrix element for the production of the
particle pairs ($i,j$) and ($k,l$) as sketched in
Fig.~\ref{fig:double-resonance} typically is of the form
\begin{align}
\mc{M}_\mr{res} = A_0 \frac{1}{\tilde{s}_{ij}} \times \frac{1}{\tilde{s}_{kl}}\,,
\end{align}
where $A_0$ denotes the finite part of the amplitude,  and the
$1/\tilde{s}_{ab}$ terms represent propagators that can go
on-shell. For each pair of external particles $(a,b)$ we introduce the invariants
$s_{ab}=(p_a+p_b)^2$ and $\tilde{s}_{ab} \equiv s_{ab} - m_{ab}^2$,
where $m_{ab}$ is the mass of the resonant particle and $p_a$, $p_b$
are the four-momenta of the considered external particles. The amplitude 
$\mc{M}_\mr{res}$ obviously diverges, if $\tilde{s}_{ij}$ and/or 
$\tilde{s}_{kl}$ approach zero.

Naively applying the regulator scheme we have used for the treatment
of single resonances, by introducing a regulator $\Gamma_\mr{reg}$,
would result in a cross section that heavily depends on the regulator.
Instead, we proceed by rewriting the resonant matrix element as  
\begin{align}
\mc{M}_\mr{res} = \frac{A_0}{\tilde{s}_{ij} + \tilde{s}_{kl}} \times
  \left(\frac{1}{\tilde{s}_{ij}} + \frac{1}{\tilde{s}_{kl}}\right)\,,
\end{align}
effectively dividing the resonance structure into two single
resonances, which can be treated separately. Since the resonance
structure of the second factor appears already at Born level, we need
to insert a regulator width only for the first term, using the same
regulator for ($i,j$) and ($k,l$). The regulated matrix element now
reads:
\begin{align}
\mc{M}_\mr{res} = \frac{A_0}{\tilde{s}_{ij} + \tilde{s}_{kl} + i
  m_{ij} \Gamma_\mr{reg} + i m_{kl} \Gamma_\mr{reg}} \times
  \left(\frac{1}{\tilde{s}_{ij}} + \frac{1}{\tilde{s}_{kl}}\right)\,.
\end{align}
In order to comply with the treatment of single resonances that
already occur at Born level, additionally the physical decay widths 
$\Gamma_{ij}^{}$ and $\Gamma_{kl}^{}$ of the resonant particles
have to be taken into account. The regulated matrix element can then be
written as
\begin{align}
\mc{M}_\mr{res} \equiv\, \mathcal{P}(\tilde{s}_{ij},\tilde{s}_{kl})\, A_0\,,
\end{align}
where $\mathcal{P}$ is the regulated propagator structure of the amplitude and is given by
\begin{align}
\mathcal{P}(\tilde{s}_{ij},\tilde{s}_{kl})\,=&\, \frac{1}{\tilde{s}_{ij} + \tilde{s}_{kl} + i
                   m_{ij} (\Gamma_\mr{reg} + \Gamma_{ij}) + i m_{kl}
                   (\Gamma_\mr{reg} + \Gamma_{kl})}\\\nonumber
 &\times \left(\frac{1}{\tilde{s}_{ij} + i m_{ij} \Gamma_{ij}} +
   \frac{1}{\tilde{s}_{kl} + i m_{kl} \Gamma_{kl}}\right)\,.
\end{align}
Following the procedure of Ref.~\cite{Baglio:2016rjx}, the calculation
of a suitable counter-term for a resonant matrix element requires the
determination of a Breit-Wigner factor that can be built by dividing
the squared propagator structure itself by the squared propagator structure taken
on-shell, where $\tilde{s}_{ij}\rightarrow 0$, $\tilde{s}_{kl}\rightarrow 0$:
\begin{align}
{\rm BW} =
  \frac{\left|\mc{P}(\tilde{s}_{ij},\tilde{s}_{kl})\right|^2}{\left|\mc{P}(0,0)\right|^2}\,.
\end{align}

Applying this procedure to the regulated doubly resonant matrix
element leads to
\begin{align}
{\rm BW} =& \Bigl[m_{ij}^2 m_{kl}^2\Gamma_{ij}^2 \Gamma_{kl}^2
            (m_{ij}(\Gamma_{ij}+\Gamma_\mr{reg})
            +m_{kl}(\Gamma_{kl}+\Gamma_\mr{reg}))^2\Bigr.\\\nonumber
& \times
  \Bigl.\left(\left(s_{ij}+s_{kl}-m_{ij}^2-m_{kl}^2\right)^2+\left(m_{ij}\Gamma_{ij}+m_{kl}\Gamma_{kl}\right)^2\right)\Bigr]/\\\nonumber 
&\Bigl[\left(\left(s_{ij}-m_{ij}^2\right)^2+m_{ij}^2
  \Gamma_{ij}^2\right)\left(\left(s_{kl}-m_{kl}^2\right)^2+m_{kl}^2
  \Gamma_{kl}^2\right)(m_{ij}\Gamma_{ij}+m_{kl}
  \Gamma_{kl})^2\Bigr.\\\nonumber
&\times\Bigl.\left(\left(s_{ij}+s_{kl}-m_{ij}^2-m_{kl}^2\right)^2+\left(m_{ij}(\Gamma_{ij}+
 \Gamma_\mr{reg})+m_{kl}(\Gamma_{kl}+\Gamma_\mr{reg})\right)^2\right)\Bigr]\,,  
\end{align}
which finally allows us to formulate the counter-term for doubly
resonant matrix elements, 
\begin{align}
\left| \mc{M}_\mr{res}^{\rm
CT}(\Gamma_{\rm reg})\right|^2 =& \,
\Theta\left(\hat{s}-(m_{ij}+m_{kl})^2\right)
\Theta\left(m_{ij}-m_{i}-m_{j}\right)\Theta\left(m_{kl}-m_{k}-m_{l}\right)\non\\
&\times {\rm BW}\times\left| \mc{M}_\mr{res}(\Gamma_\mr{reg})\right|^2_\mr{remapped} \,.
\label{eq:onshell2}
\end{align}
Here $\hat{s}$ denotes the partonic center-of mass energy, and
$m_{a}$ ($a = (ij),(kl))$ the mass of an on-shell particle. 
The momenta entering $\mc{M}_\mr{res}$ in the on-shell counter-term are to be
remapped to the on-shell kinematics of the resonant particles,
c.f.~Ref.~\cite{GoncalvesNetto:2012yt}.

Similar to the singly resonant case, the cross section corresponding to
the on-shell subtracted resonant terms in the doubly resonant case is
obtained by subtracting the counter-term from the resonant matrix element and
summing over all possible resonant channels:
\begin{align}
\sigma_{\rm real}^{\rm OS} &= \sum\limits_{\mr{res}}\int d\Phi_{4}
\left[\left| \mc{M}_\mr{res}(\Gamma_\mr{reg})\right|^2 -
                             \mc{J}_{\mr{res}}\left| \mc{M}_\mr{res}^{\rm
                             CT}(\Gamma_\mr{reg})\right|^2\right]\,,
\end{align}
where the Jacobian factor $\mc{J}_{\mr{res}}^{}$ reads
\begin{align}
\mc{J}_{\mr{res}}^{} &= \frac{s_{ij}s_{kl}}{m_{ij}^2m_{kl}^2}
  \frac{\lambda^{1/2}(\hat{s},m_{ij}^2,m_{kl}^2)
  \,\lambda^{1/2}(m_{ij}^2,m_{i}^2,m_{j}^2)\,\lambda^{1/2}(m_{kl}^2,m_{k}^2,m_{l}^2)}
{\lambda^{1/2}(\hat{s},s_{ij},s_{kl})\,\lambda^{1/2}(s_{ij},m_{i}^2,m_{j}^2)\,\lambda^{1/2}(s_{kl},m_{k}^2,m_{l}^2)} \nonumber\\
&\times \frac{(\sqrt{\hat{s}}-m_{ij})^2-(m_k+m_l)^2}{(\sqrt{\hat{s}}-\sqrt{s_{ij}})^2-(m_k+m_l)^2}\, . 
\end{align}
Here, $\lambda$ denotes the usual Kaellen-function $\lambda(x,y,z) =
x^2+y^2+z^2-2(xy+yz+zx)$, and $d\Phi_{4}^{}$ is the full
$2\to 4$ phase-space element.
For the actual evaluation of  $\sigma_{\rm real}^{\rm OS}$ in the
\POWHEGBOX{}, we have devised a routine allowing for a mapping of the
phase-space according to a process-specific resonance structure.
Because the real emission amplitudes require a special treatment for
the removal of  the doubly on-shell contributions, obtaining them with
a default amplitude generator was not  possible. We thus have used a
modified version of {\tt FeynArts~3.9}~\cite{Hahn:2000kx} and {\tt 
FormCalc~9.4}~\cite{Hahn:1998yk} for the generation of the
real-emission amplitudes and the identification of their resonance
structure. 

\begin{figure}[t]
\bec
\includegraphics[width=0.48\textwidth]{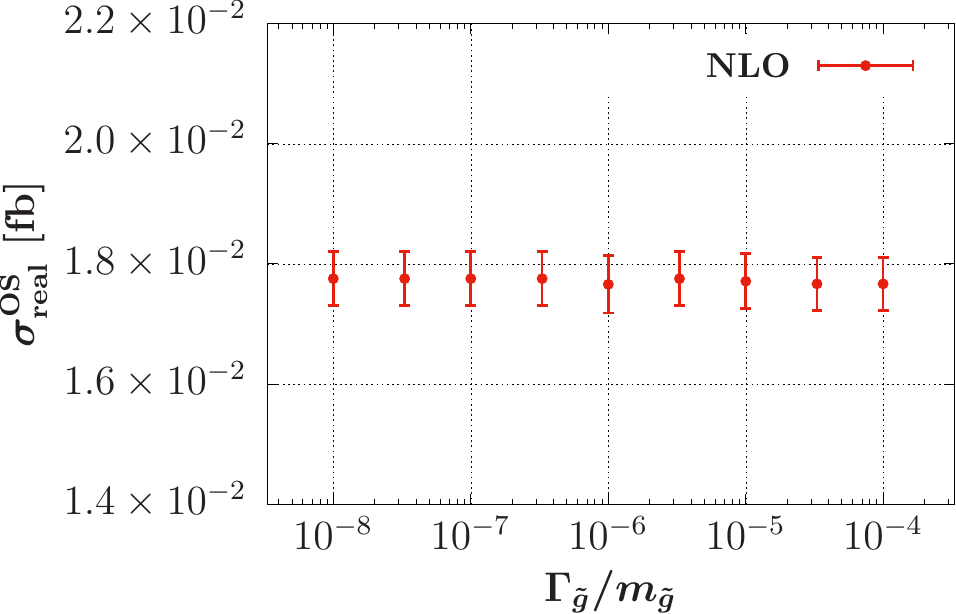}\hspace{2mm}
\includegraphics[width=0.48\textwidth]{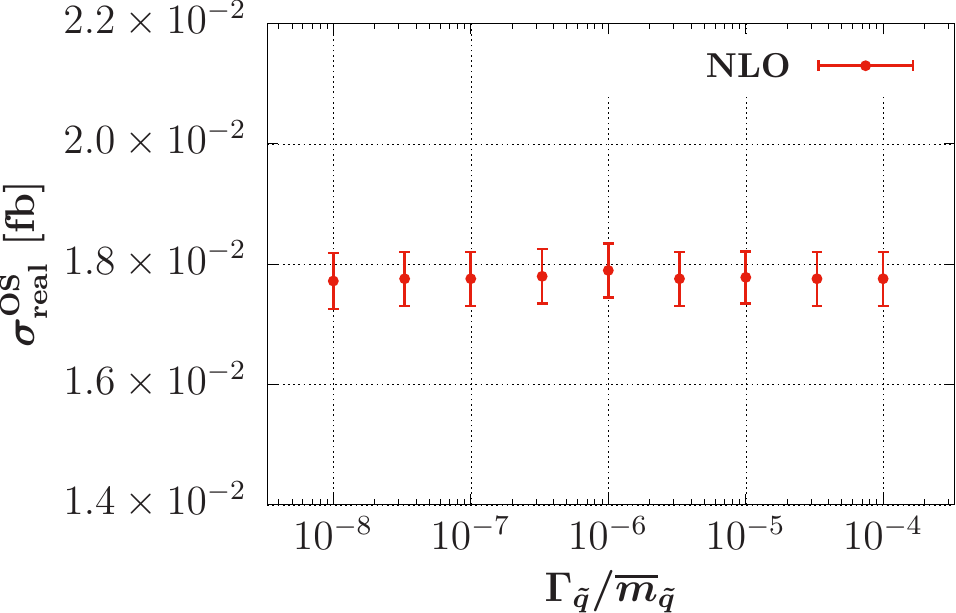}
\caption{Dependence of the real on-shell contribution $\sigma_{\rm
    real}^{\rm OS}$ to the cross section for the process $pp\to
  \neu_1^0\neu_1^+\jet$ with $\sqrt{s}=14$~TeV on the technical
  regulator $\Gamma_{\tilde{g}}$ (left) and $\Gamma_{\tilde{q}}$
  (right).}
\label{fig:reg-dep}
\eec
\end{figure}

In order to verify the validity of our implementation, we have
performed a number of checks. First, we have tested that, after the
subtraction of on-shell resonances, for collinear momentum
configurations real-emission and IR subtraction terms approach each
other.
Second, we have found that the dependence of our predictions for
weakino-pair plus jet production cross sections on the technical regulator
$\Gamma_\mr{reg}$ is negligible. Figure~\ref{fig:reg-dep} illustrates
the regulator dependence of the on-shell resonant part of
neutralino-pair production in association with a jet for a SUSY benchmark
point that features squarks heavy enough to on-shell decay into a 
neutralino and a quark.
Using the average of the four squark masses of the first generation, 
$\overline{m}_{\tilde{q}} = 1.67$~TeV, and the gluino mass 
$m_{\tilde{g}} = 1.78$~TeV, we find that in the range
$\Gamma_{\tilde{q}}/\overline{m}_{\tilde{q}} = 10^{-8}\;\mr{to}\;10^{-4}$ 
and $\Gamma_{\tilde{g}}/m_{\tilde{g}} = 10^{-8}\;\mr{to}\;10^{-4}$, the 
dependence of the cross section on the regulator is entirely
negligible, thus confirming the stability of the applied on-shell
subtraction procedure.
Additionally, we have checked that the proposed on-shell subtraction method 
preserves gauge invariance for the processes under consideration.

Finally, we have computed inclusive cross sections at LO and NLO accuracy
for the pair production of the lightest neutralino in association with
a jet in the setup of Ref.~\cite{Cullen:2012eh} and found good
agreement with the published results within the range of one
percent. It is important to mention that the calculation described in
Ref.~\cite{Cullen:2012eh} does not include on-shell resonances, which
we accounted for when comparing the results. Contributions of on-shell
diagrams may be small for many SUSY parameter points and can be
neglected in such cases, but they are essential for a consistent
calculation in the most general case,  as some allowed corners of the
SUSY spectrum may lead to on-shell resonances.

%

%
\section{Phenomenological results}
\label{sec:pheno}

A public release of the code for the production of a weakino pair with
an additional identified jet will be made available in the framework of
the \POWHEGBOX{} via the project website
{\tt \url{http://powhegbox.mib.infn.it}}. In this section we intend to
highlight representative phenomenological results in order to
demonstrate the capabilities of our code.

\subsection{Choice of the spectrum and input parameters}
In order to calculate the physical SUSY spectrum and to obtain a SUSY
Les Houches Accord (SLHA) file~\cite{Skands:2003cj,Allanach:2008qq}
 as input to our code, we chose a parameter point in the
framework of the pMSSM10~\cite{deVries:2015hva} that is still allowed
by current experimental limits on SUSY. The pMSSM10 is defined at the
SUSY scale $M_{\rm SUSY} =
\sqrt{m_{\tilde{t}_1^{}}^{}m_{\tilde{t}_2^{}}^{}}$, where
$\tilde{t}_1$ and $\tilde{t}_2$ are the two stop mass-eigenstates, 
with ten soft SUSY breaking parameters, namely: the
gaugino masses $M_{1}$, $M_{2}$, $M_{3}$, the first- and
second-generation squark masses that are taken to be equal,
$m_{\tilde{q}_1^{}}^{}=m_{\tilde{q}_2^{}}^{}$, the third-generation
squark mass $m_{\tilde{q}_3^{}}^{}$, a common slepton mass for the
three generations $m_{\tilde \ell}^{}$, a common trilinear mixing
parameter $A$ for the three generations, the Higgsino mass parameter
$\mu$, the pseudo-scalar mass $M_A^{}$, and $\tan\beta$, the ratio
between the two vacuum expectation values of the Higgs fields. Left-
and right-handed sfermion soft breaking masses are taken to be equal.

We have chosen a scenario in which the LSP is the lightest neutralino, 
with a reasonably low mass so that the production
cross section is not too low. This has in particular lead to the
following values of the ten parameters highlighted above,
\bea
  M_1^{} = -120~\text{GeV},\ \ \ \ 
  M_2^{} = 160~\text{GeV},\ \ \ \
  M_3^{} = 1.70~\text{TeV},\nonumber
\eea
\vspace{-8mm}
\bea
  m_{\tilde{q}_1^{}}^{} = 1.79~\text{TeV},\ \ \ \ 
  m_{\tilde{q}_3^{}}^{} = 1.30~\text{TeV},\ \ \ \ 
  m_{\tilde{\ell}}^{} = 740~\text{GeV},\ \ \ \
  A = 1.863~\text{TeV},\nonumber
\eea
\vspace{-8mm}
\bea
  \mu = 190~\text{GeV},\ \ \ \ 
  M_A^{} = 1.35~\text{TeV},\ \ \ \
  \tan\beta = 35.
\eea
We have used the {\tt SoftSUSY 4.0} program~\cite{Allanach:2001kg} for
the calculation of the spectrum, and the {\tt SDECAY 1.3}
program~\cite{Muhlleitner:2003vg} for the calculation of the relevant
decay widths and branching fractions to obtain the SLHA input file for
our code in the \POWHEGBOX. Our electroweak input parameters are the
$Z$ boson mass, $m_Z^{} = 91.1876$~GeV, the electromagnetic coupling
constant $\alpha^{-1}_{}(M_Z^{}) = 127.934$, and the Fermi constant,
$G_F^{} = 1.16637\times 10^{-5}_{}$ GeV$^{-2}_{}$. The resulting
neutralino masses are
\bea
m_{\neu_1^0} = 111.9~\GeV, \ \ \ \
m_{\neu_2^0} = 129.3~\GeV, \ \ \ \
m_{\neu_3^0} = 211.7~\GeV, \ \ \ \
m_{\neu_4^0} = 245.6~\GeV,\eea
and the chargino masses are
\bea
m_{\neu_1^\pm} = 130.9~\GeV, \ \ \ \
m_{\neu_2^\pm} = 249.1~\GeV\,.
\eea
The squark masses are equal for the first and second generation (and
for up- and down-type squarks), but different for the third
generation. They read
\bea
m_{\dsql/\ssql} = m_{\usql/\csql} = 1.836~\TeV, \ \ 
m_{\dsqr/\ssqr} = m_{\usqr/\csqr} = 1.833~\TeV\,,\nonumber
\eea
\vspace{-8mm}
\bea
m_{\bsql} = 1.330~\TeV, \ \ 
m_{\bsqr} = 1.346~\TeV, \ \ 
m_{\tsql} = 1.229~\TeV, \ \ 
m_{\tsqr} = 1.423~\TeV\,.
\eea

\subsection{Cross sections and distributions at the LHC}
We consider proton-proton collisions at the LHC with a center-of-mass
energy of $\sqrt{s} = 14$~TeV. For the parton distribution functions
(PDFs) of the proton we use for both the LO and NLO calculation the
'PDF4LHC15\_nlo\_nf4\_30' PDF set~\cite{Butterworth:2015oua} with four
active flavors as implemented in the LHAPDF
library~\cite{Buckley:2014ana} with ID $= 92000$. We set the
renormalization and factorizations scales, $\mur$ and $\muf$, to be
proportional to the sum of the masses of the weakinos $\neu_A$ and
$\neu_B$ produced in the specific process under consideration,
$\mur=\muf=\xi\mu_0 $ with $\mu_0 = m_{\neu_A}+m_{\neu_B}$, while the
scale parameter $\xi$ is chosen to be one unless specifically stated
otherwise. The fixed-order results are combined with the parton-shower
program \PYTHIA{}~6.4.25~\cite{Sjostrand:2006za} in which QED
radiation, underlying event, and hadronization effects are switched
off throughout. Partons arising from the real-emission contributions
of the NLO-QCD calculation or from the parton shower are recombined
into jets according to the anti-$k_T$ algorithm~\cite{Cacciari:2008gp}
as implemented in the {\tt FASTJET} package~\cite{Cacciari:2011ma}
with $R=0.4$ and $\left|\eta^\mr{jet}\right|<4.5$.

In the following we discuss phenomenological results for neutralino-pair 
production in association with a hard jet, $p p \to \neu_1^0
\neu_1^0 j$, in two different setups. For the first one
only basic jet selection cuts are applied on the transverse momentum
$p_T^{\mr{jet}_{1}}$ and rapidity $\eta^{\mr{jet}_{1}}$ of the hardest
identified jet,
\begin{align}
\label{eq:Jcuts}
p_T^{\mr{jet}_{1}}  > 20~\GeV\,, \quad
|\eta^{\mr{jet}_{1}}| < 4.5\,,
\end{align}
which results in a cross section of $1.37~\fb$ at NLO accuracy.
The second set of cuts is inspired by experimental mono-jet searches
\cite{Aaboud:2017nhr,Aad:2015zva,EXOT:2015005}. Here, in addition to a
severe cut on the hardest jet, an additional  cut on the missing
transverse momentum $p_T^\mr{miss}$ is imposed,
\begin{align}
\label{eq:Mcuts}
p_T^\mr{miss} > 100~\GeV\,, \quad
p_T^{\mr{jet}_{1}} > 80~\GeV\,, \quad
|\eta^{\mr{jet}_{1}}| < 2.8\,. 
\end{align}

The  missing transverse momentum is computed from the negative sum of
detectable jets with a transverse momentum $p_T^\mr{jet} \geq 30$~GeV
and $|\eta^{\mr{jet}}| < 2.8$, similar to what is done in the
experimental analyses. As the sum of the transverse momenta of the
final-state particles needs to add to zero, this is effectively
equivalent to the sum over the non-detected particles, more
specifically the LSP and the softer jets.

\begin{figure}[t!]
\bec
\includegraphics[width=0.48\textwidth]{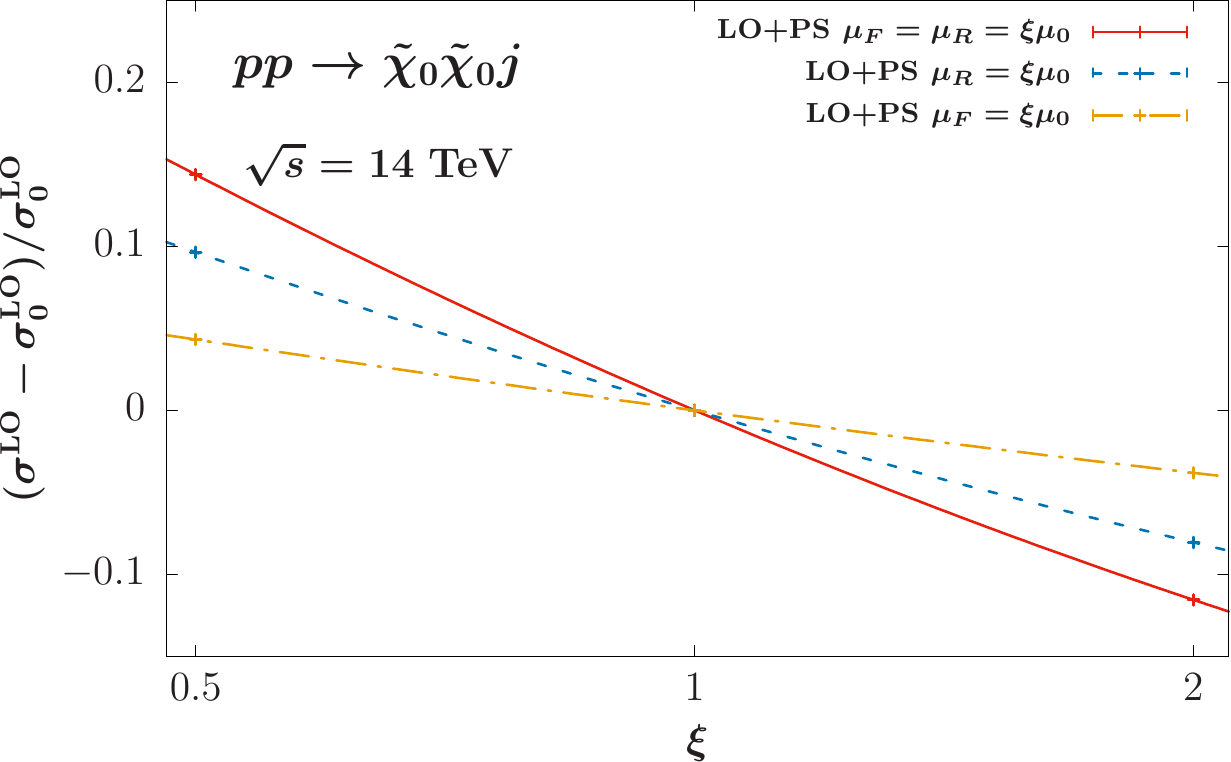}\hspace{2mm}
\includegraphics[width=0.48\textwidth]{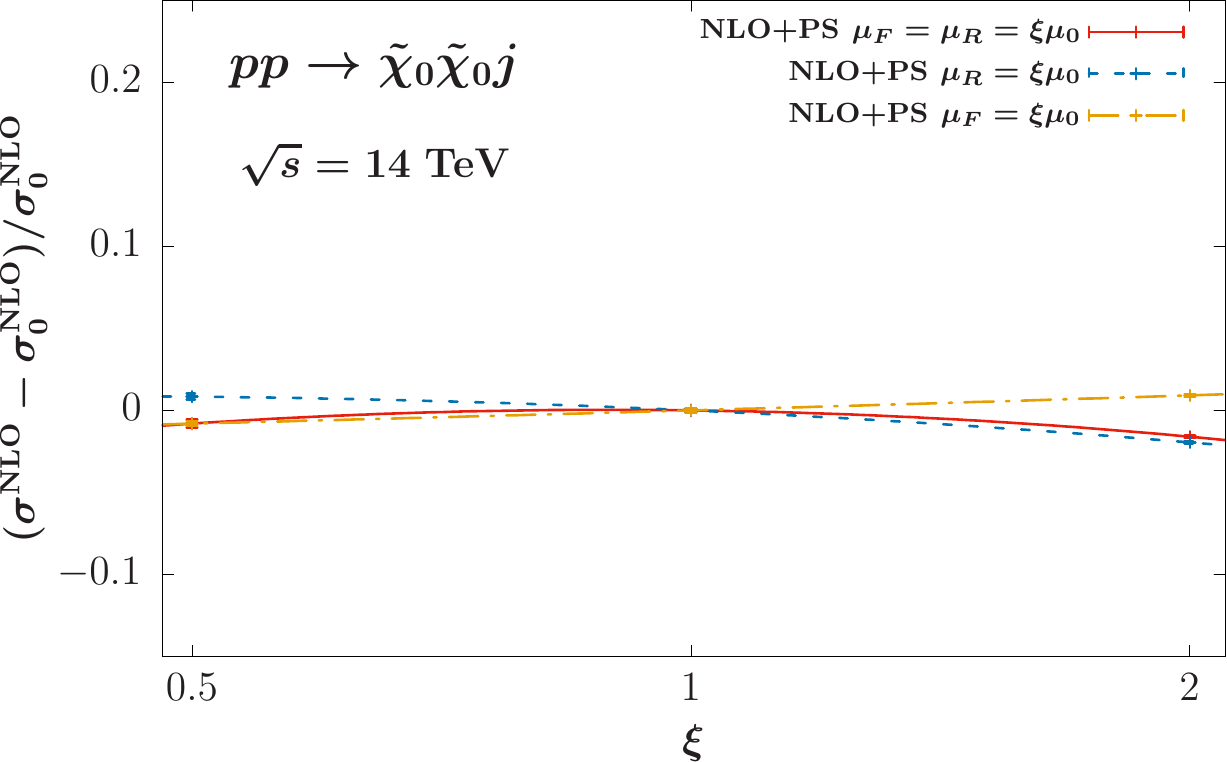}
\caption{Dependence of the inclusive cross section for the process $pp\to\neu_1^0\neu_1^0\jet$
  with $\sqrt{s}=14$~TeV within the cuts of Eq.~(\ref{eq:Jcuts}) on 
  the factorization and renormalization scales.
  The curves show the deviation, $(\sigma - \sigma_0)/\sigma_0$, 
  from the respective  LO (left) or NLO (right) cross section,  
  $\sigma_0 = \sigma(\mur = \muf = \mu_0^{})$, 
  as a function of the scale parameter $\xi$, for three different cases: $\mur=\muf=\xi \mu_0$
  (solid red line), $\mur=\xi \mu_0,\,\muf=\mu_0$ (dashed blue line), and
  $\mur=\mu_0,\,\muf=\xi \mu_0$ (dot-dashed yellow line). In each case, $\mu_0=2\,m_{\neu_1^0}$.}
\label{fig:scale-dep}
\eec
\end{figure}

Figure~\ref{fig:scale-dep} illustrates the dependence of 
the LO and NLO cross sections within the cuts of 
Eq.~(\ref{eq:Jcuts}) on the scale parameter $\xi$. To quantify the 
theoretical uncertainties which emerge from the unphysical 
renormalization and factorization scales, we have varied $\mu_R$ and 
$\mu_F$ in the range $0.5\mu_0$ to $2\mu_0$ around the default 
choice $\mu_0 \equiv 2 m_{\tilde{\chi}_1^0}$. Since the LO cross 
section is already dependent on the strong coupling constant 
$\alpha_s$, neutralino-pair production in association with a jet 
depends not only on $\mu_F$ via the parton distribution functions of 
the scattering protons, but also on the renormalization scale 
$\mu_R$ entering the running of the strong coupling 
constant $\alpha_s^{}$. At NLO, additional $\mu_R$ dependence 
occurs in the form of loop diagrams. However, in the considered range 
$0.5\mu_0$ to $2\mu_0$ the NLO cross section changes by only about $3$\%, 
whereas the LO cross section changes by up to $14$\%. This indicates that 
the perturbative expansion is stable, and the scale uncertainty at NLO 
is reduced remarkably compared to the LO calculation.

\begin{figure}[t!]
\bec
\includegraphics[width=0.48\textwidth]{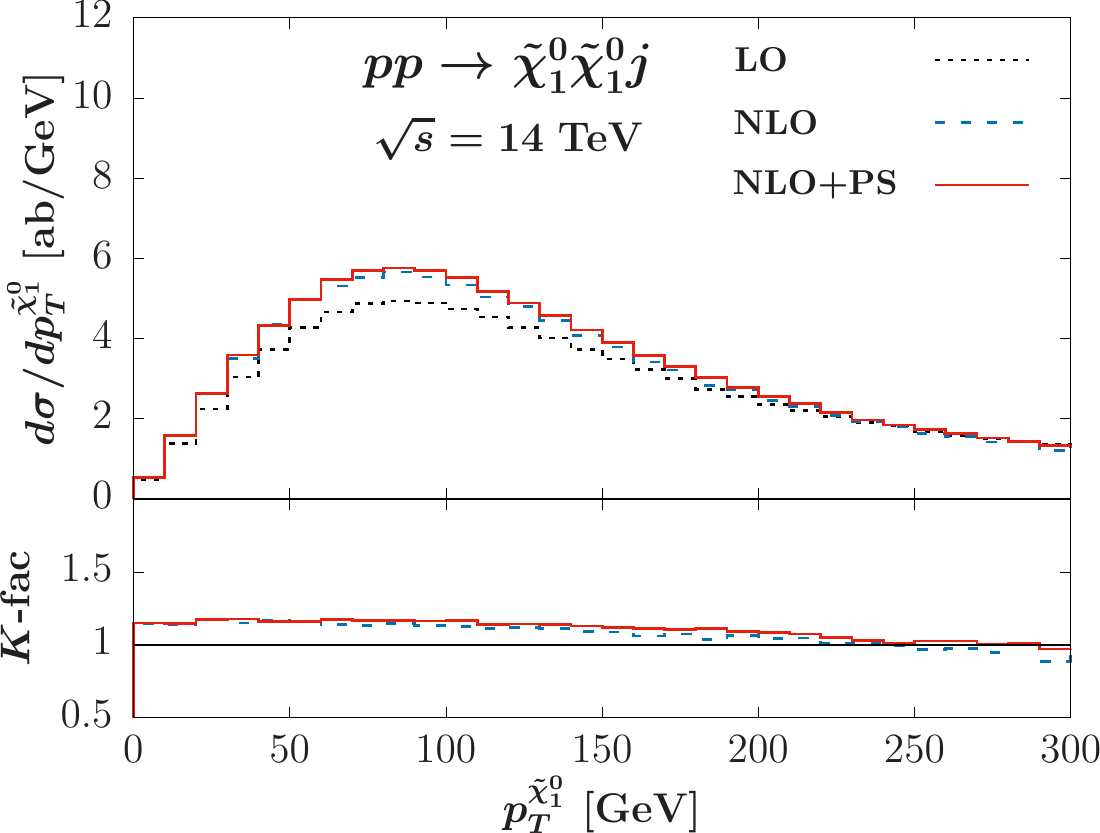}\hspace{2mm}
\includegraphics[width=0.48\textwidth]{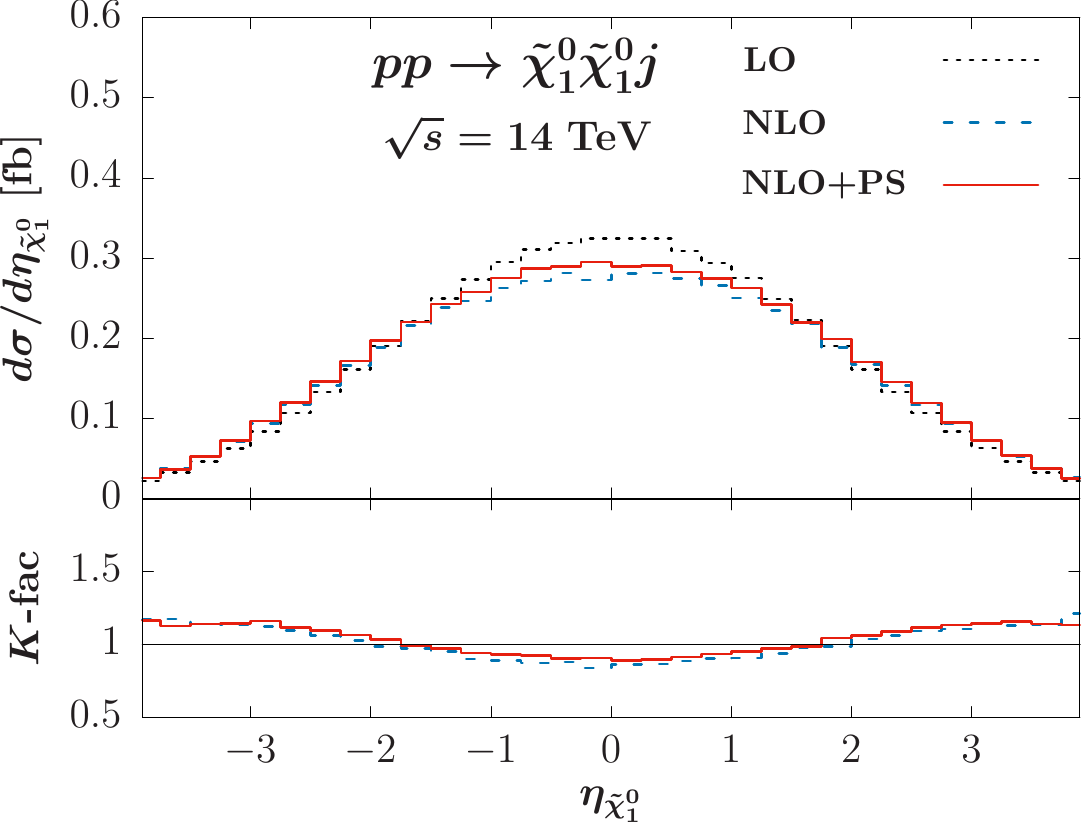}
\caption{Average transverse-momentum $p_T^{\neu_1^0}$ (left)
  and pseudo-rapidity $\eta_{\neu_1^0}^{}$
  distributions (right) of the two neutralinos in the process $pp\to
  \neu_1^0\neu_1^0\jet$ at LO (dotted black lines), NLO (dashed blue
  lines), and \NLOPS{} (solid red lines) within the cuts of
  Eq.~(\ref{eq:Jcuts}) for our default setup.}
\label{fig:ptAetaA}
\eec
\end{figure}

\begin{figure}[t!]
\bec
\includegraphics[width=0.48\textwidth]{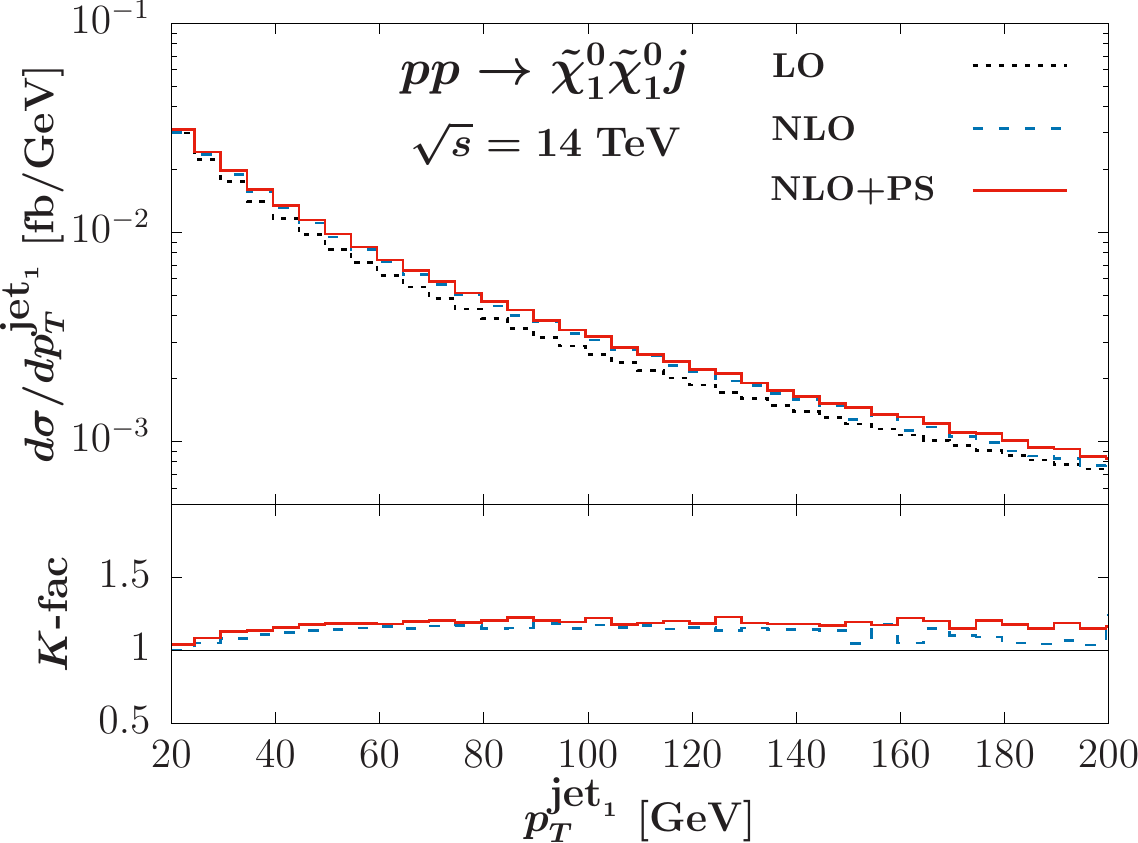}\hspace{2mm}
\includegraphics[width=0.48\textwidth]{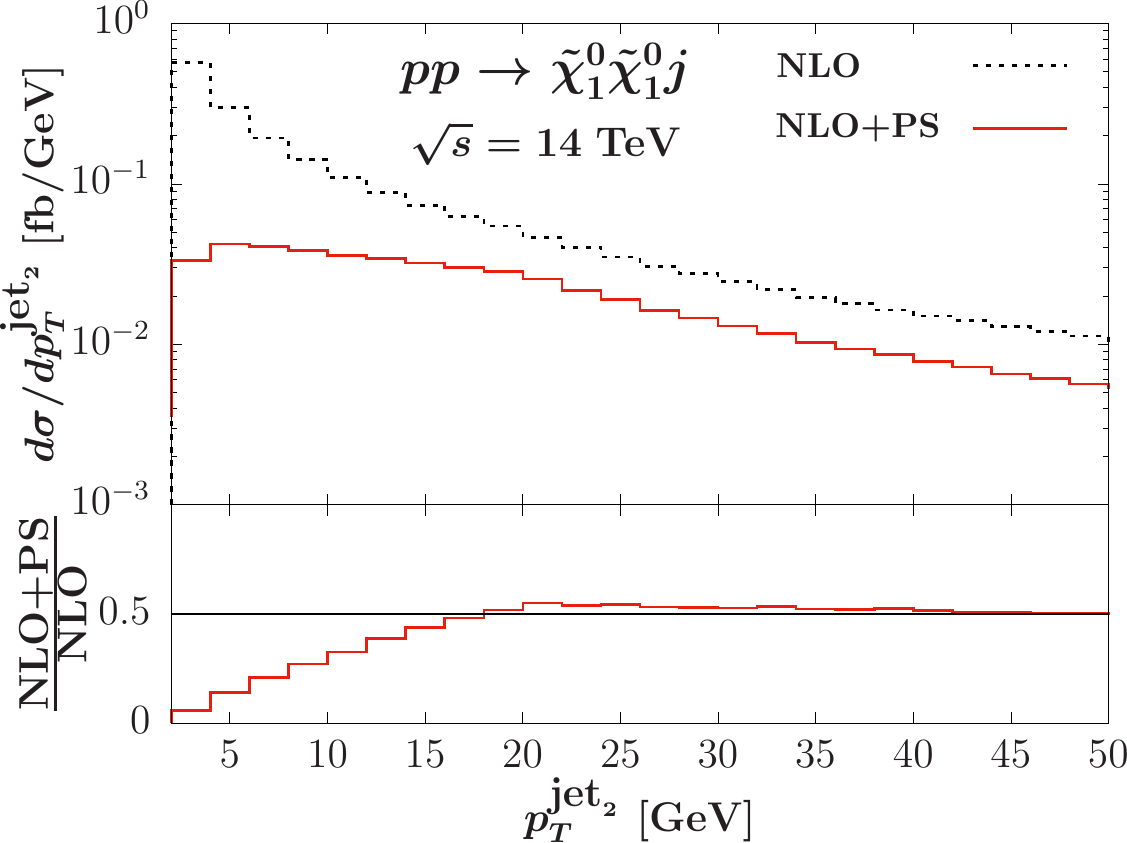}
\caption{Transverse-momentum distribution of the hardest jet (left)
  and the second hardest jet (right) for the process $pp\to
  \neu_1^0\neu_1^0\jet$ at LO (dotted black lines), NLO (dashed blue
  lines), and \NLOPS{} (solid red lines) for our default setup, after
  the cuts of Eq.~(\ref{eq:Jcuts}) are applied.}
\label{fig:ptJ1ptJ2}
\eec
\end{figure}

Characteristic distributions of the two neutralinos for the inclusive
setup of Eq.~(\ref{eq:Jcuts}) are shown in Fig.~\ref{fig:ptAetaA}.
%
%
%
%
In each case, the NLO prediction is significantly different from the
respective LO curve. Adding the parton shower results in a 
small increase of the associated cross section. In the
transverse-momentum distribution of the two neutralinos the NLO
corrections are largest in the low-$p_T$ range, amounting to almost
$15$\% while they decrease to less than $5$\% in the tail of the
distribution. The parton shower increases the distribution uniformly
over the entire range by about $5$\%.
In the average pseudo-rapidity distribution of the neutralinos, the corrections are negative in the central rapidity
region, amounting to $-15$\%. Towards larger values of $\eta_{\neu_1^{0}}^{}$, the corrections are positive and increase the LO result by $15$\%.
Similar to the transverse-momentum distribution of the two neutralinos, the parton shower increases the $\eta_{\neu_1^{0}}^{}$-distribution by $5$\% over the entire range.
The transverse-momentum distribution of the hardest jet, depicted in
Fig.~\ref{fig:ptJ1ptJ2}, exhibits a behavior slightly different than
of the neutralinos. NLO corrections are small in the bulk, but increase
up to $15$\% in the tail of the distribution. Parton-shower effects
modify the fixed-order NLO results by an additional $5$\%.
Obviously, a constant $K$--factor would not account for these effects in each of these distributions.

While the distributions of the hardest jet can be described with full
NLO accuracy by our calculation, the second-hardest jet is accounted
for only by the real-emission contributions of the NLO-QCD
corrections, and thus effectively only described with LO
accuracy. Since no cuts are imposed on sub-leading jets in the setup
we consider, it would be expected that the soft and collinear
configurations dominate the behavior of such jets. This expectation is
nicely confirmed for the transverse-momentum distribution of the
second-hardest jet in Fig.~\ref{fig:ptJ1ptJ2}.
%
%
%
%
The curve associated with the NLO calculation increases rapidly
towards small values of $p_T^\mr{jet_2}$ and exhibits a large negative
entry in the lowest bin. The Sudakov factor of the \NLOPS{}
implementation tames this increase.

\begin{figure}[t!]
\bec
\includegraphics[width=0.47\textwidth]{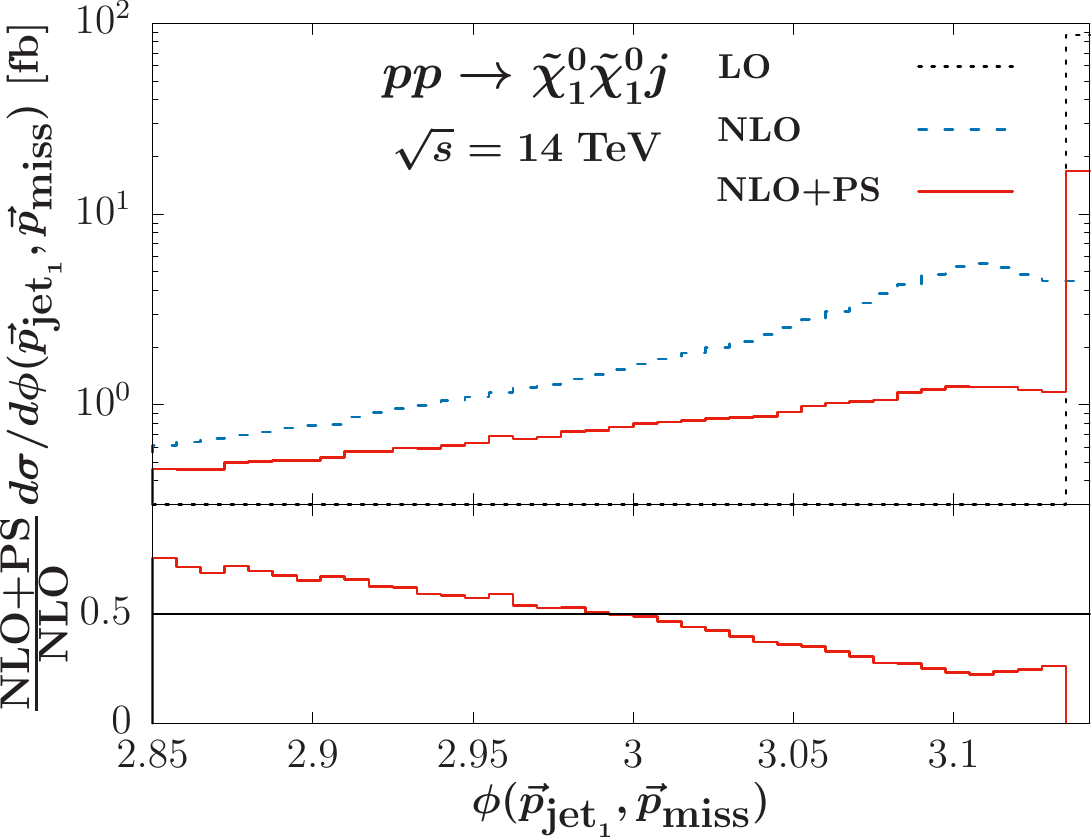}\hspace{2mm}
\includegraphics[width=0.49\textwidth]{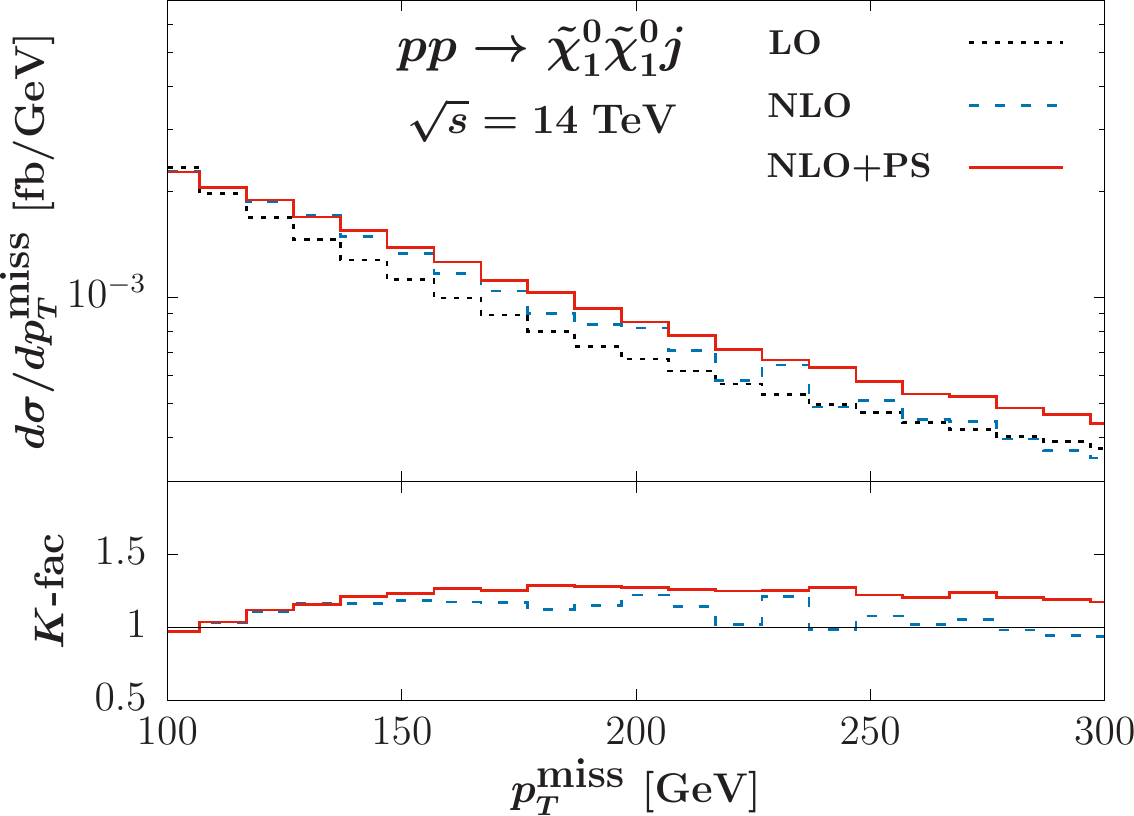}
\caption{Angular separation $\phi$ between
the missing momentum
  $\vec{p}_\mr{miss}$ and the hardest jet momentum
  $\vec{p}_{\mr{jet}_{1}}$ vectors (left) and missing transverse
  momentum $p_T^{\rm miss}$ (right) for the process $pp\to \neu_1^0\neu_1^0\jet$ at LO
  (dotted black lines), NLO (dashed blue lines) and \NLOPS{} (solid red lines), 
  after the cuts of Eqs.~(\ref{eq:Mcuts}) are applied.}
\label{fig:phiptMiss}
\eec
\end{figure}

Experimental searches for particles that cannot be directly identified
in a default detector, such as neutralinos or, more generically,
massive DM candidate particles, often rely on mono-jets. The tell-tale
signature of such events consists in a hard jet that recoils off the
system comprised by the heavy particles, accompanied by large missing 
transverse energy. To quantitatively account for such signatures  an 
accurate description of the hard jet accompanying the heavy-particle 
system is of paramount importance. 
For the production of a neutralino pair in association with a hard jet
that can give rise to a  mono-jet signature we  consider a scenario
inspired by experimental searches with the cuts of
Eq.~(\ref{eq:Mcuts}). Figure~\ref{fig:phiptMiss}
%
%
%
%
illustrates the features of two characteristic distributions for such
a scenario: The angular separation
$\phi(\vec{p}_\mr{jet_1^{}},\vec{p}_\mr{miss})$ of the hardest jet and 
of the missing momentum $\vec{p}_\mr{miss}$;  and the  transverse
component of the missing momentum, $p_T^\mr{miss}$. At LO, the hard
jet is produced back-to-back with the heavy neutralino system,
resulting in an angular separation of $180^\circ$. At NLO, this
back-to-back configuration can be altered by radiation effects due to
real parton emission. Even more radiation occurs at the \NLOPS{} level,
resulting in an additional smearing of the LO distribution and affecting
in particular the last bin corresponding to $\phi = 180^\circ$.
The $p_T^\mr{miss}$ distribution experiences a pronounced increase
from LO to NLO and \NLOPS{}. While the NLO corrections are
moderate in the bulk, they amount to more than $15$\% beyond about
$150$~GeV. A description of this behavior by a constant K~factor would
clearly fail. The $p_T^\mr{miss}$ distribution is increased uniformly
by another $10$\% from NLO to \NLOPS{} by the parton-shower.  
 
\begin{figure}[t!]
\bec
\includegraphics[width=0.485\textwidth]{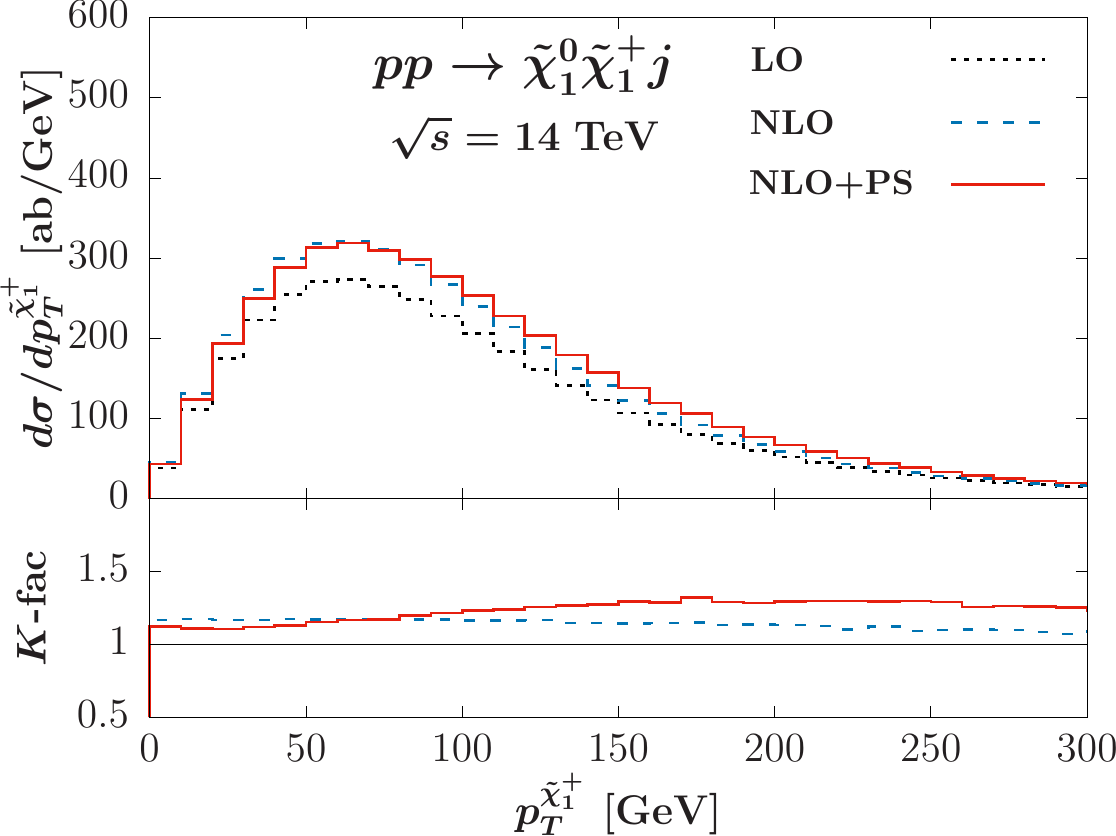}\hspace{2mm}
\includegraphics[width=0.475\textwidth]{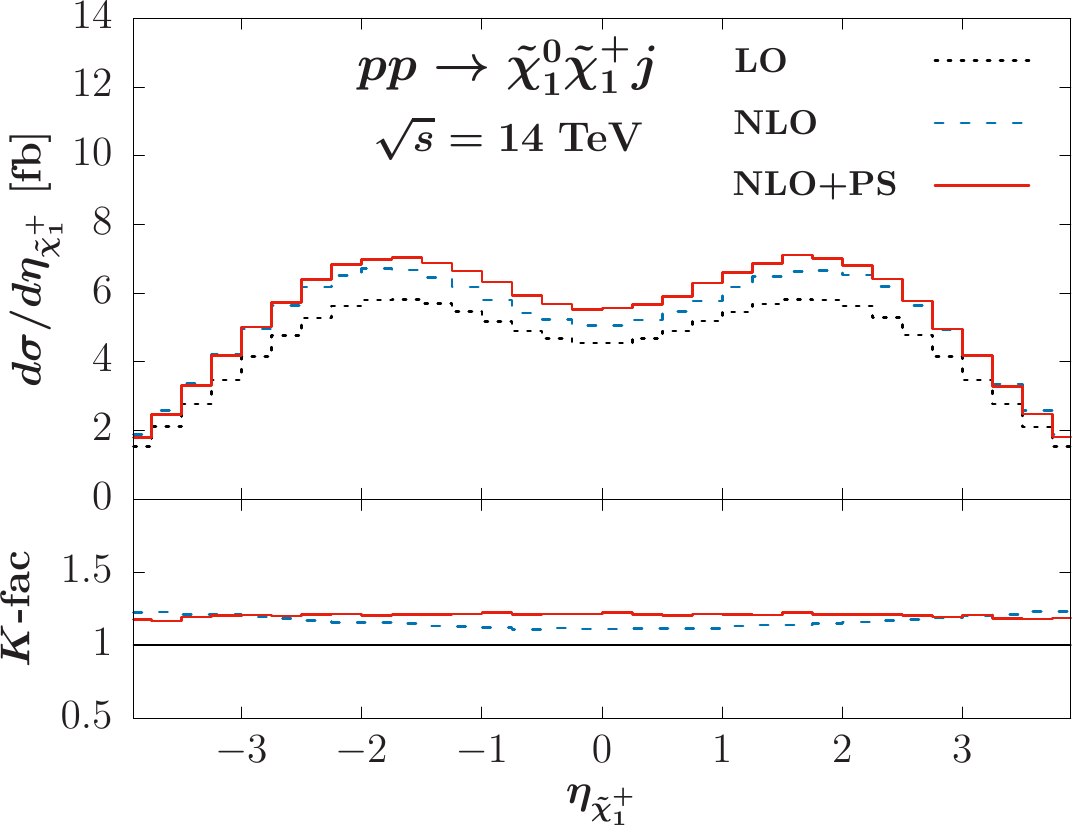}
\caption{Transverse-momentum (left) and pseudo-rapidity distribution
  (right) of the chargino in the process $pp\to
  \neu_1^0\cha_1^+\jet$ at LO (dotted black lines), NLO (dashed blue
  lines), and \NLOPS{} (solid red lines) within the cuts of Eq.~(\ref{eq:Jcuts}) for our default setup.}
\label{fig:ptBetaB}
\eec
\end{figure}

While a major motivation for the investigation of
weakino-pair production processes at hadron colliders is the search
for Dark Matter, this class of reactions is interesting also per se
for the study of SUSY interactions. The code package we have developed
thus not only allows for simulations of neutralino-pair production in
association with a hard jet, but also for production processes
involving various combinations of charginos and neutralinos. To
demonstrate this feature of our work,  we present in
Fig.~\ref{fig:ptBetaB} representative distributions of the chargino
produced in the reaction $pp\to \neu_1^0\cha_1^+\jet$ after the
inclusive selection cuts of Eq.~(\ref{eq:Jcuts}) are applied.
We find that the NLO (SUSY-)QCD corrections to the transverse-momentum
and pseudo-rapidity distributions of the chargino are similar in size
to those for the related distributions in the case of neutralino-pair
production (c.f.~Fig.~\ref{fig:ptAetaA}).
While we refrain from such a study in this work, we would like to
point out that event files produced with our code in the default Les
Houches format can easily be processed with public Monte-Carlo
generators such as {\tt PYTHIA} to simulate decays of the
supersymmetric particles. In that case, not only distributions of the
charginos, but also of their decay products can be simulated. 
%

%
\section{Conclusions}
\label{sec:conc}
In this paper we have presented the implementation of  weakino-pair production 
in association with an identified jet at a hadron collider 
in the framework of the \POWHEGBOX{}. The newly developed code allows for the
calculation of the NLO SUSY-QCD corrections for the hard production
process, and provides an interface to parton-shower programs such as
\PYTHIA{} via the \POWHEG{} method. The program can process SLHA files
obtained with an external spectrum calculator for the computation of a
specific SUSY parameter point in the context of the MSSM. A
generalized method to subtract on-shell resonances consistently 
in a $2\to 3$ process has been formulated for the first time
which could be helpful for many of other processes where the
convergence of perturbation theory is spoiled by double on-shell
resonances in the real corrections.

To illustrate the capabilities of the developed code package,
we have discussed phenomenological features of a few selected
processes focusing on theoretical uncertainties and the impact of
parton shower effects on experimentally accessible observables. We
have found that, in accordance with previous results reported in the
literature, generally NLO corrections have a significant impact on
production rates and reduce the scale uncertainty of the theoretical calculation.
Parton-shower effects are small for weakino
distributions, but are significant for jet observables, as
expected. Thus, our work is of immediate relevance for mono-jet 
searches for Dark Matter at the LHC in the framework of the MSSM.
%

%
\acknowledgments
We are very grateful to Thomas Hahn for generous help with {\tt
  FormCalc} and {\tt FeynArts}. We also thank Marco Stratmann and
Christoph Borschensky for valuable comments and discussions. 
We furthermore would like to thank the referee of our manuscript for feedback that helped to improve our article.
This work
has been supported in part by the Institutional Strategy of the
University of T\"ubingen (DFG, ZUK 63), by the DFG Grant JA 1954/1, by
the Carl Zeiss Foundation, and by the German Academic Scholarship
Foundation (Studienstiftung des deutschen Volkes). J.B. also
acknowledges the support in the last stages of this work from his
Durham Senior Research Fellowship COFUNDed between Durham University
and the European Union under grant agreement number 609412. This work
was performed thanks to the support of the state of
Baden-W\"{u}rttemberg through bwHPC and the German Research Foundation
(DFG) through grant no INST 39/963-1 FUGG. The Feynman diagrams of
this paper have been drawn with the program {\tt JaxoDraw
  2.0}~\cite{Binosi:2003yf,Binosi:2008ig}.

%

\bibliography{weakino_jet}{}
\bibliographystyle{JHEP}

\end{document}